\documentclass[10pt]{article}
\usepackage{graphicx}

\def\Journal#1#2#3#4{{#1} {\bf #2}, #3 (#4)}

\title{Spectra of identified hadrons with the ALICE detector in pp and Pb--Pb collisions at the LHC
}
\author{Roberto Preghenella (for the ALICE Collaboration)\\
  \small Centro Studi e Ricerche e Museo Storico della Fisica ``Enrico Fermi'',
  Rome, Italy\\
  \small Dipartimento di Fisica dell'Universit\`a and Sezione INFN, Bologna,
  Italy\\
  \small E-mail: {\tt preghenella@bo.infn.it}
}
\date{}

\begin{document}
\maketitle
\begin{abstract}
The measurement of identified charged-hadron production at mid-rapidity ($|y| < 0.5$) performed
with the ALICE experiment at the LHC is presented for pp collisions at $\sqrt{s}$ = 900 GeV and 7 TeV
and for Pb--Pb collisions at $\sqrt{s_{NN}}$ = 2.76 TeV. Transverse momentum spectra of $\pi^{\pm}$, K$^{\pm}$, p, $\rm \bar{p}$ and multi-strange baryons are measured over a wide momentum range using the $dE/dx$, the time-of-flight and topological
 particle-identification techniques. In this report, the particle-identification detectors and techniques, 
as well as the achieved performance, are shortly reviewed. Proton-proton results on particle production yields, 
spectral shapes and particle ratios are presented as a function of the collision energy and compared to previous 
experiments and commonly-used Monte Carlo models. Particle spectra, yields and ratios in Pb--Pb are measured 
as a function of the collision centrality and the results are compared with published RHIC data in Au--Au 
collisions at $\sqrt{s_{NN}}$ = 200 GeV and predictions for the LHC.
\end{abstract}

\section{Introduction}\label{sec:introduction}
ALICE (A Large Ion Collider Experiment) is a general-purpose heavy-ion
experiment at the CERN LHC (Large Hadron Collider) aimed at studying the
physics of strongly-interacting matter and the quark--gluon plasma. A unique
design has been adopted for the ALICE detector to fulfill tracking and
particle-identification requirements~\cite{ref:ALICEperf}. Thanks to these features the experiment
is able to identify hadrons in a wide momentum range by combining
different detecting systems and techniques, as discussed in 
Section~\ref{sec:pid}. Results on
hadron spectra and yields at mid-rapidity are presented in
Section~\ref{sec:ppresults} for pp collisions at $\sqrt{s}$~=~900~GeV and
7~TeV and in Section~\ref{sec:pbpbresults} for Pb--Pb collisions at
$\sqrt{s_{\rm NN}}~=~\rm~2.76~TeV$.

\section{Particle identification}\label{sec:pid}

In this section the particle-identification (PID) detectors relevant for the
analyses presented in this paper are briefly discussed, namely the \emph{Inner
  Tracking System} (ITS), 
the \emph{Time-Projection Chamber} (TPC) and the \emph{Time-Of-Flight} detector
(TOF). A detailed review of the ALICE detector and of its PID capabilities can
be found in~\cite{ref:ALICEperf}. The ITS is a six-layer silicon detector located at radii between 4 and 43
cm. Four of the six layers provide $dE/dx$ measurements and are used for
hadron identification in the non-relativistic ($1/\beta^2$)
region. Moreover, using the ITS as a standalone tracker enables one to
reconstruct and identify low-momentum hadrons not reaching the main tracking
systems. The TPC is the main central-barrel tracking detector of ALICE and
provides three-dimensional hit information and specific energy-loss
measurements with up to 159 samples. With the measured particle momentum and
$\langle dE/dx \rangle$ the hadron type can be determined by comparing the
measurements against the Bethe-Bloch expectation. The TOF detector is a
large-area array of Multigap Resistive Plate Chambers (MRPC) and covers the central
pseudorapidity region ($\left| \eta \right| <$~0.9, full azimuth). Hadron
identification is performed by matching momentum and trajectory-length
measurements performed by the tracking system with the time-of-flight
information provided by the TOF system. The total time-of-flight resolution is
about 85 ps in Pb--Pb collisions and it is determined by the time resolution
of the detector itself and by the start-time resolution.

The transverse momentum spectra of primary $\pi^{\pm}$, $K^{\pm}$, $p$ and
$\bar{p}$ are measured at mid-rapidity ($\left|y\right|~<~0.5$) combining the
techniques and detectors described above. Primary hadrons
are defined as prompt particles produced in the collision and all decay
products, except products from weak decay of strange particles. The
contribution from the feed-down of weakly-decaying particles to $\pi^{\pm}$,
$p$ and $\bar{p}$ and from protons from material are subtracted by fitting the
data using Monte Carlo templates of the DCA\footnote{Distance of Closest
  Approach to the reconstructed primary vertex.} distributions. Particles can
also be identified in ALICE via their characteristics decay 
topology or invariant mass fits. This, combined with the direct identification
of the decay daughters allows to reconstruct weakly-decaying particles and
hadronic resonances with an improved signal-to-background ratio.

\section{Results in pp collisions}\label{sec:ppresults}

\begin{figure}[t]
  \centering
  \begin{minipage}[c]{0.49\linewidth}
    \centering
    \includegraphics[viewport=10 0 517 366, width=\textwidth]{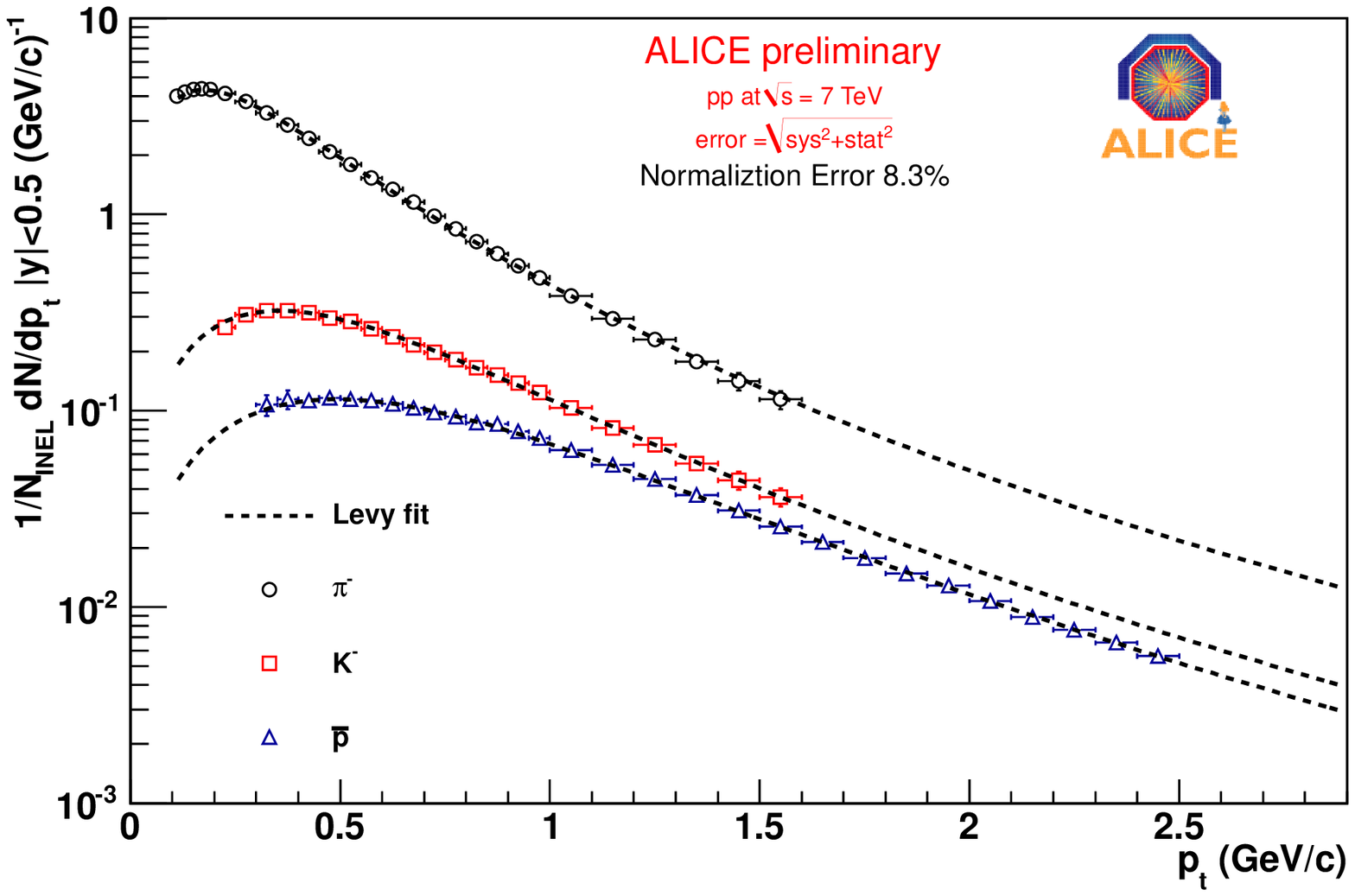}
  \end{minipage}
  \hfill
  \begin{minipage}[c]{0.49\linewidth}
    \centering
    \includegraphics[viewport=10 0 517 384, width=\linewidth]{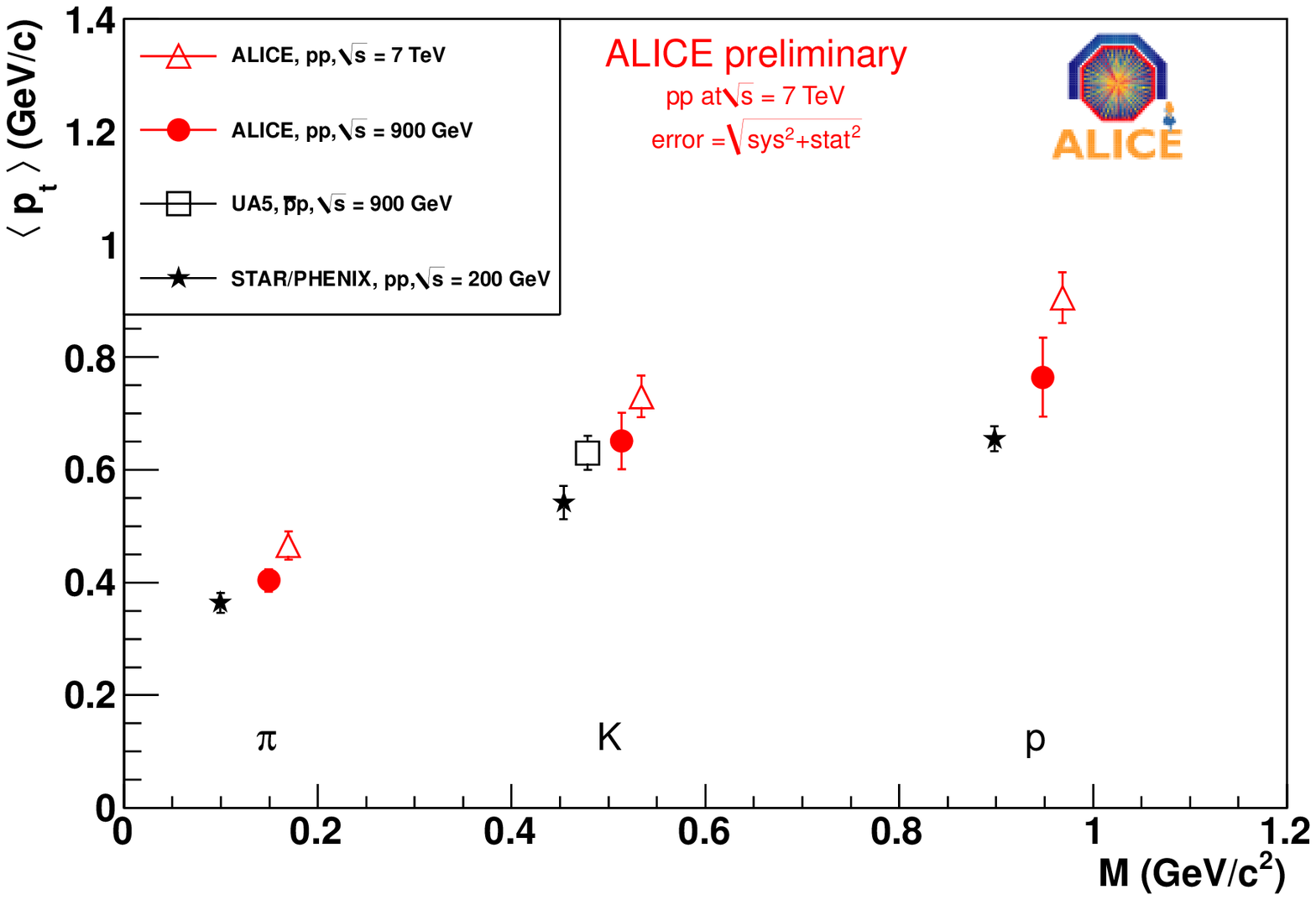}
  \end{minipage}
  \caption{Transverse momentum spectra of primary $\pi^{-}$, $K^{-}$,
    $\bar{p}$ and corresponding fits in pp collisions at
    $\sqrt{s}$~=~7~TeV (left). Mean $p_{T}$ as function of the hadron mass (right).}
    \label{fig:ppspectrameanptpp}
\end{figure}

\begin{figure}[t]
  \centering
  \begin{minipage}[c]{0.49\linewidth}
    \centering
    \includegraphics[viewport=10 0 537 366, width=\textwidth]{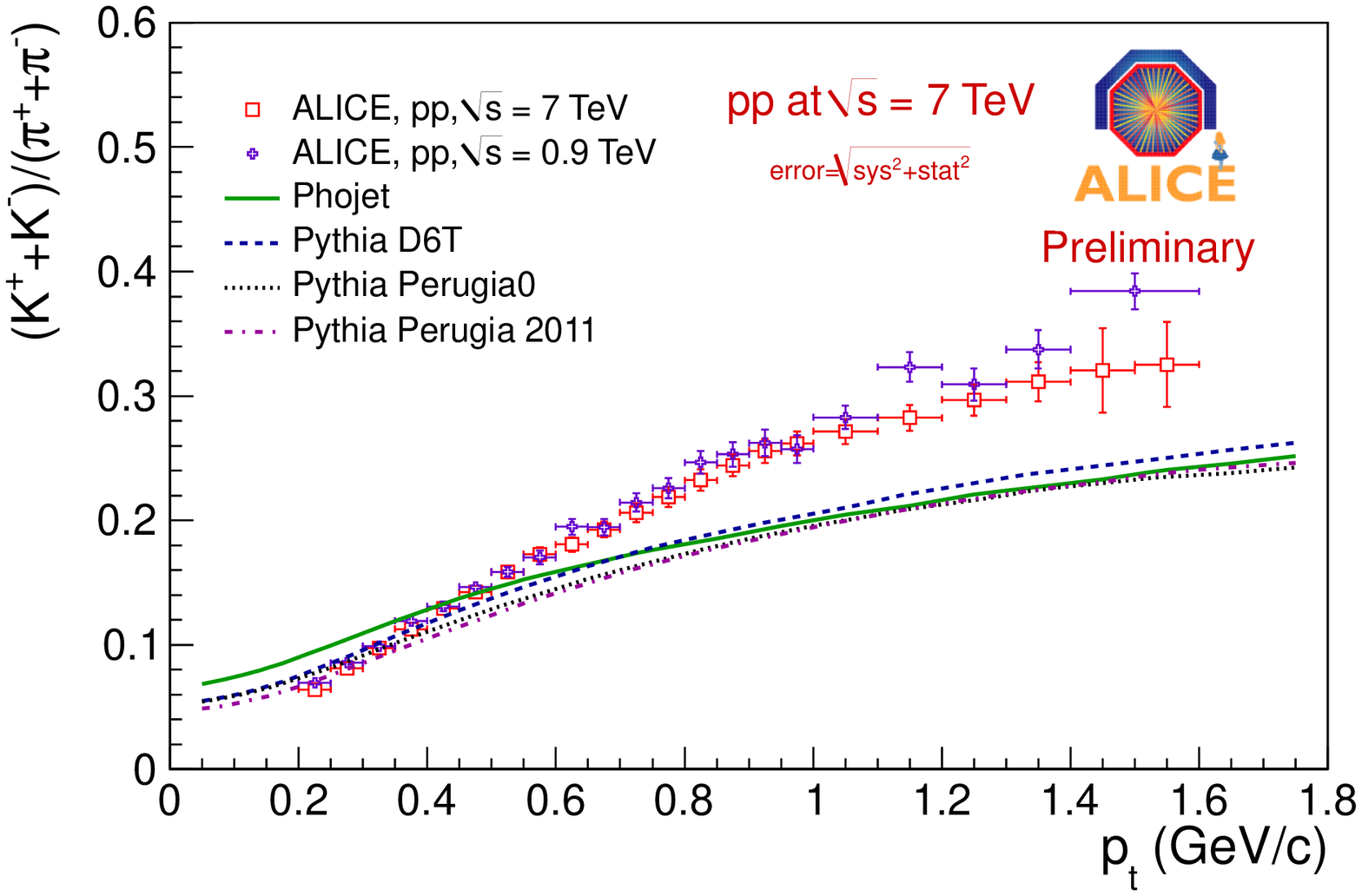}
  \end{minipage}
  \hfill
  \begin{minipage}[c]{0.49\linewidth}
    \centering
    \includegraphics[viewport=10 0 537 366, width=\linewidth]{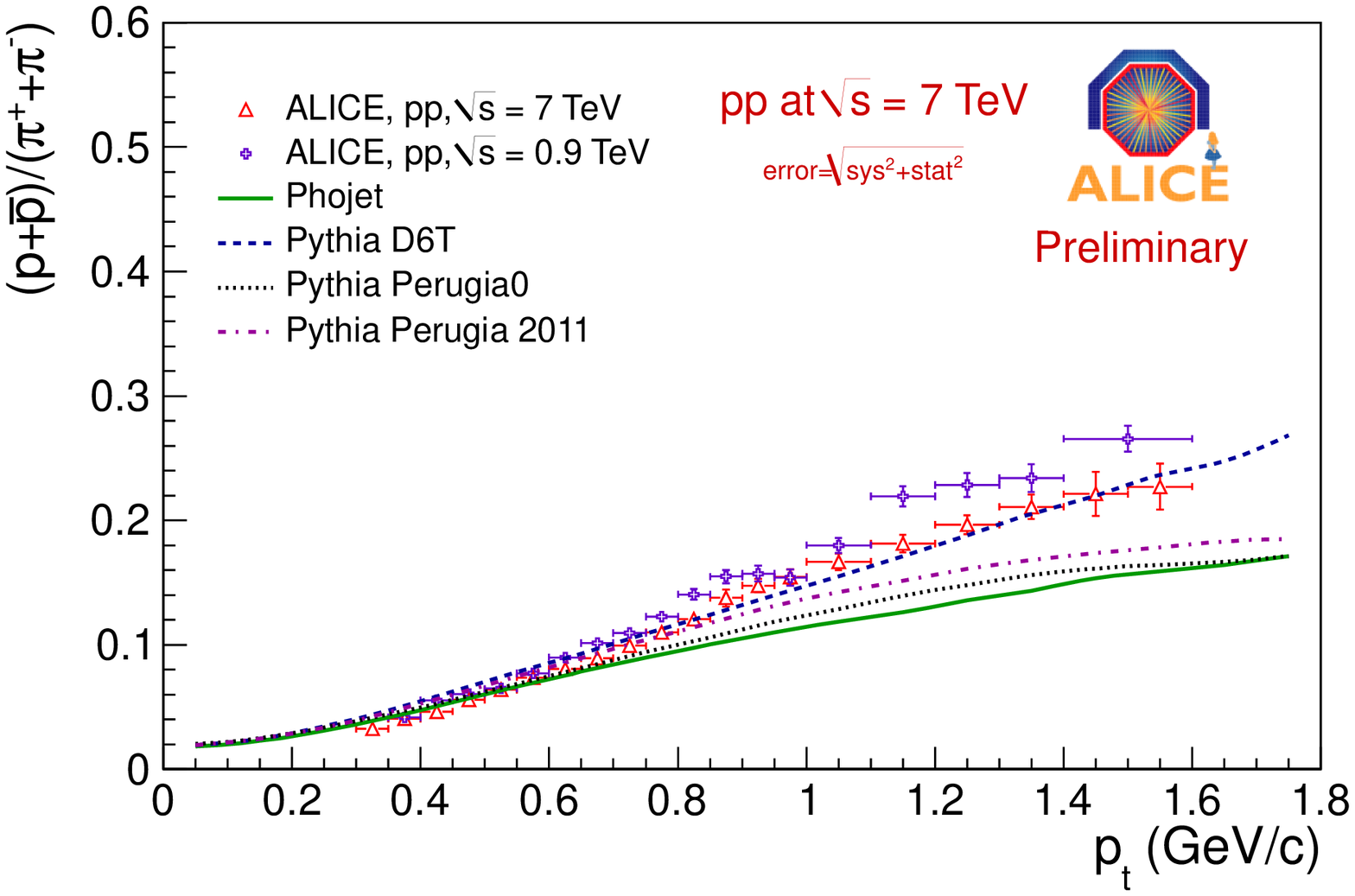}
  \end{minipage}
  \caption{$K/\pi$ (left) and $p/\pi$ (right) production ratios as a function
    of $p_{T}$ measured in pp collisions at $\sqrt{s}$~=~900~GeV and 7~TeV
    compared to various Monte Carlo models at $\sqrt{s}$~=~7~TeV.}
  \label{fig:ratiopp}
\end{figure}

\begin{figure}[t]
  \centering
  \begin{minipage}[c]{0.48\linewidth}
    \centering
    \includegraphics[viewport=10 0 517 366, width=\textwidth]{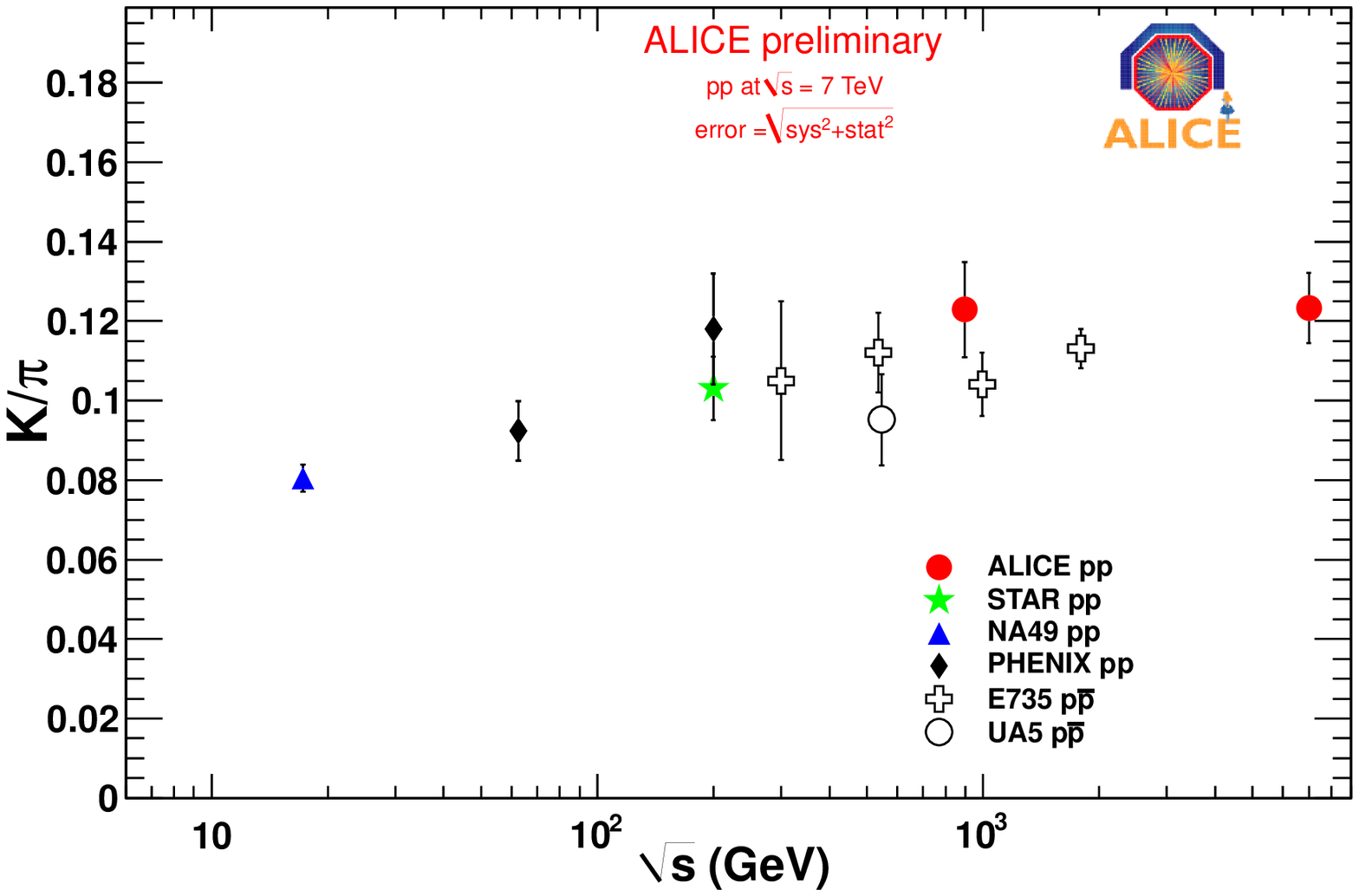}
  \end{minipage}
  \hfill
  \begin{minipage}[c]{0.48\linewidth}
    \centering
    \includegraphics[viewport=10 0 517 366, width=\linewidth]{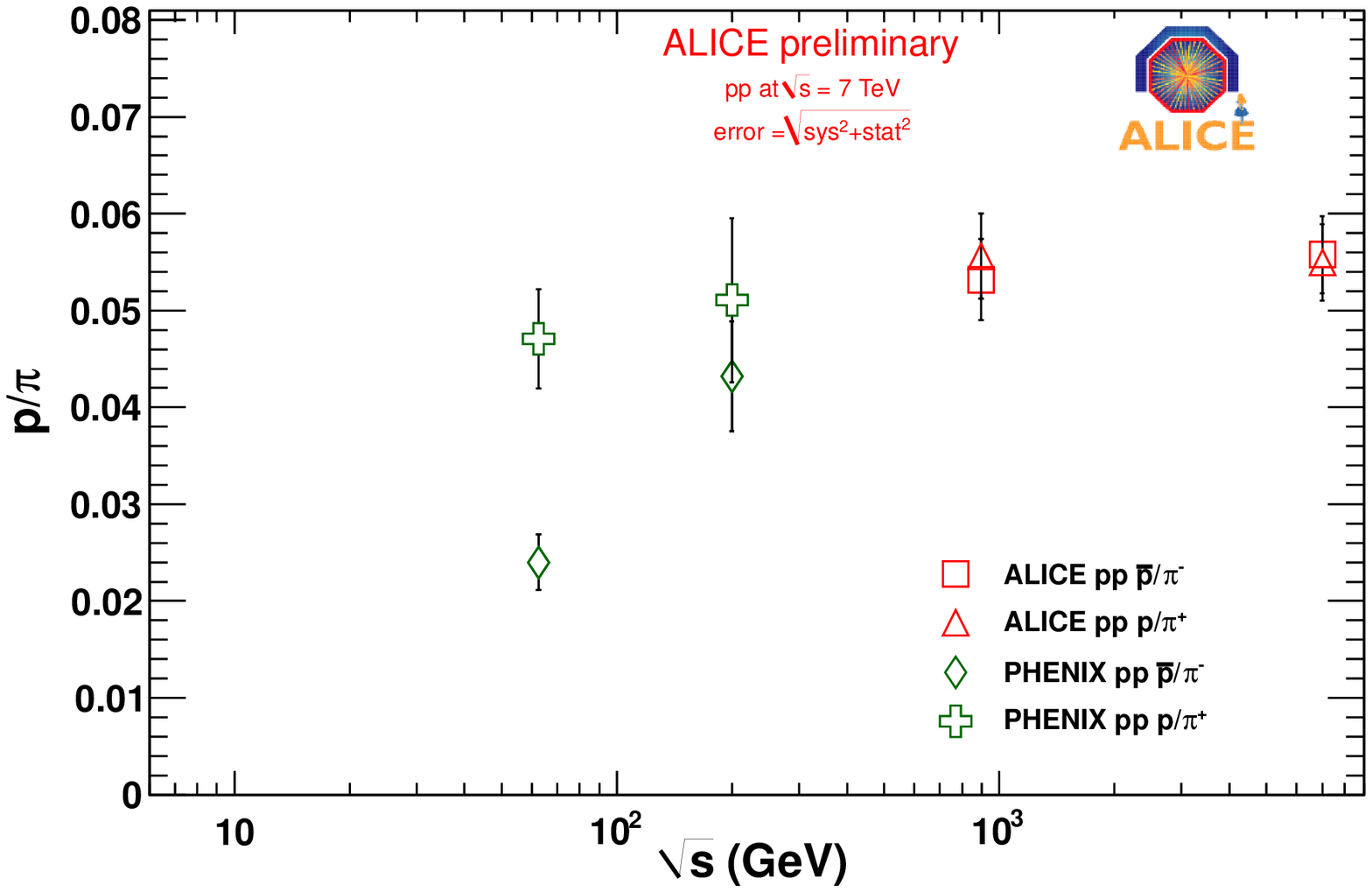}
  \end{minipage}
  \caption{$K/\pi$ (left) and $p/\pi$ (right) integrated production ratios as a
    function of $\sqrt{s}$.}
  \label{fig:ratiopps}
\end{figure}

The transverse momentum spectra of primary $\pi^{\pm}$, $K^{\pm}$, $p$ and
$\bar{p}$ are measured in minimum-bias pp collisions at
$\sqrt{s}$~=~900~GeV and 7~TeV. The measurements performed at
$\sqrt{s}$~=~900~GeV and the details of the analysis are already published
in~\cite{ref:spectra900}. The results of the analysis performed at
$\sqrt{s}$~=~7~TeV together with the corresponding fits to the data are shown
in Figure~\ref{fig:ppspectrameanptpp} (left) for negative hadrons. The
L\'evy-Tsallis~\cite{ref:tsallis} parameterization provides a good description of the spectral
shapes at both energies and allows to extrapolate the spectra outside the
measured $p_{T}$ range to compute integrated yields $dN/dy$ and average transverse
momenta $\langle p_{T} \rangle$. Figure~\ref{fig:ppspectrameanptpp} (right) compares $\langle p_{T}
\rangle$ at different energies and colliding systems which is observed to
rise with increasing $\sqrt{s}$. The ratios $K/\pi$ and $p/\pi$ as a function of $p_{T}$ are shown in
Figure~\ref{fig:ratiopp} comparing measurements at $\sqrt{s}$~=~900~GeV and
7~TeV. Both particle ratios do not show evident energy dependence. A
comparison with Monte Carlo generators shows that $K/\pi$ ratio is
underestimated at high-$p_{T}$ by recent PYTHIA tunes. The same
holds for $p/\pi$ ratio, though a better agreement with the data is observed
for PYTHIA D6T. Figure~\ref{fig:ratiopps} shows the integrated production ratios $K/\pi$ and
$p/\pi$ as a function of $\sqrt{s}$. They are observed to be rather
independent of the collision energy from 900~GeV to 7~TeV. Moreover, for
these energies there are no difference between $p/\pi^{+}$ and
$\bar{p}/\pi^{-}$, hence the baryon/antibaryon asymmetry vanishes at LHC
energies as already reported in~\cite{ref:panos} leading to a constant value of about
0.05.

\begin{figure}[t]
  \centering
  \begin{minipage}[c]{0.48\linewidth}
    \centering
    \includegraphics[width=\textwidth]{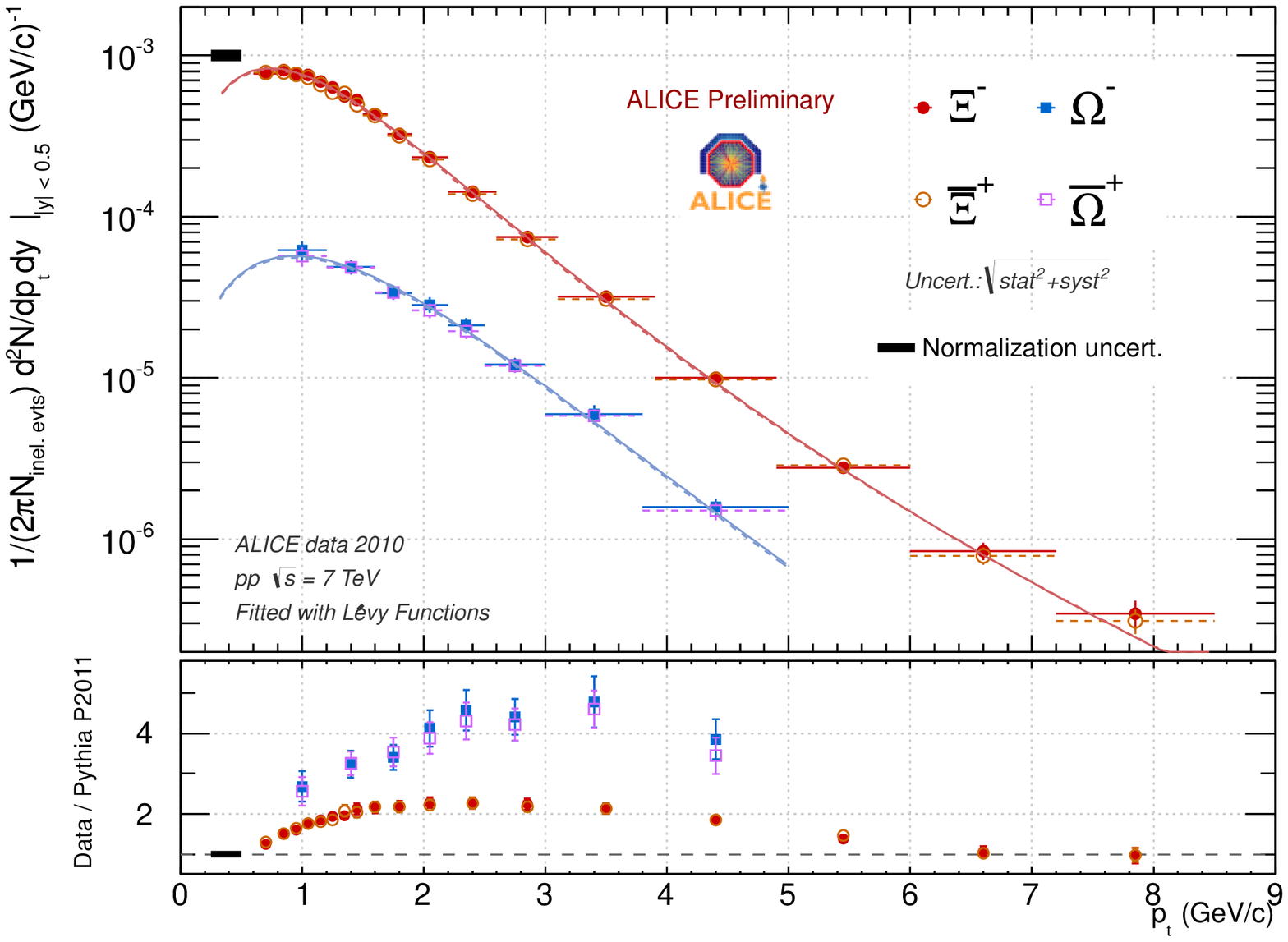}
  \end{minipage}
  \hfill
  \begin{minipage}[c]{0.48\linewidth}
    \centering
    \includegraphics[width=\linewidth]{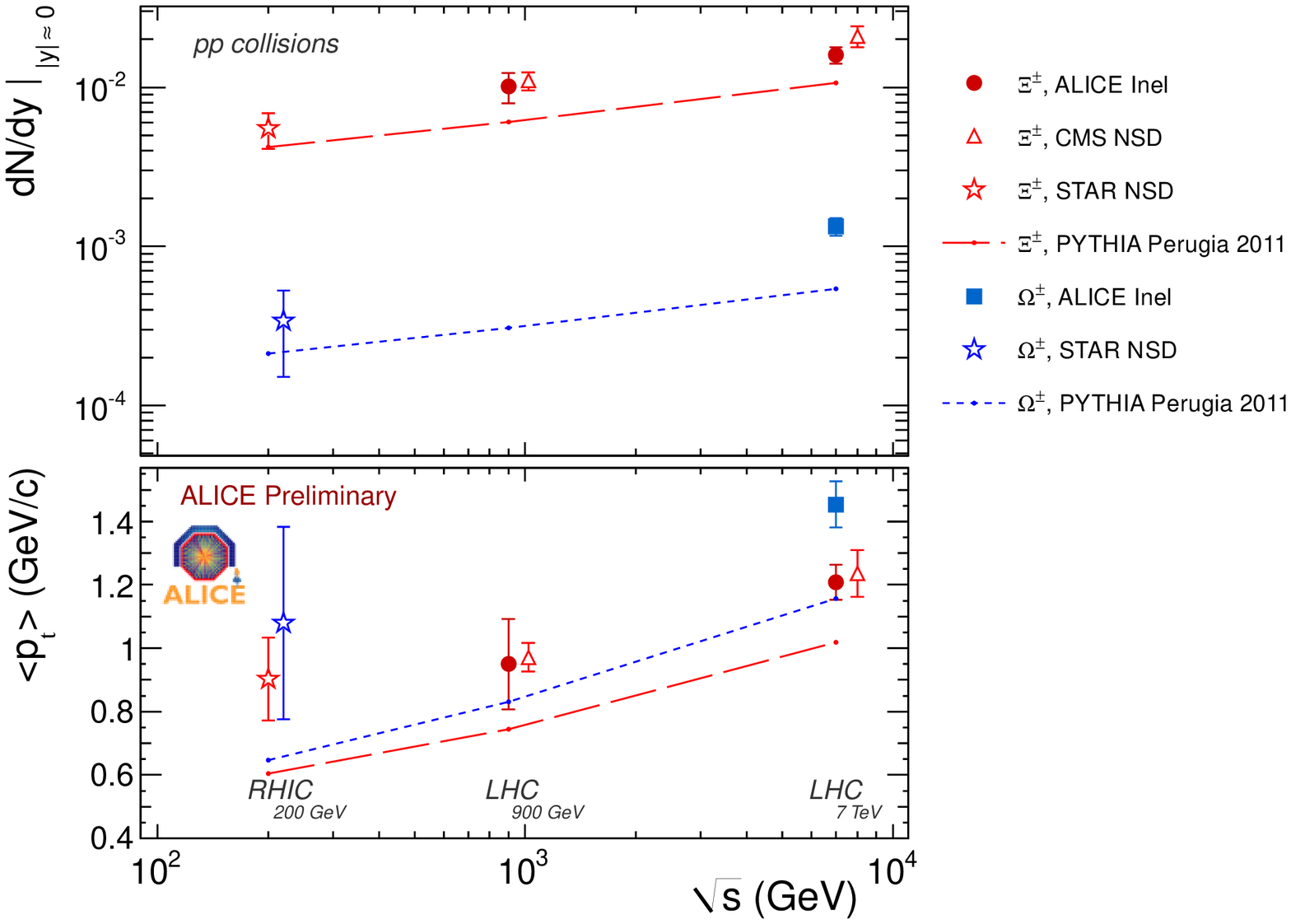}
  \end{minipage}
  \caption{$\Xi^{-}$ and $\Omega^{-}$ baryon and their antiparticle spectra (left, top panel) and the comparison to PYTHIA Perugia 2011 Monte Carlo (left, bottom panel). $dN/dy$ (right, top panel) and $\langle p_{t} \rangle$ (right, bottom panel) of $\Xi^{\pm}$ and $\Omega^{\pm}$ as a function of collision energy.}
  \label{fig:multipp}
\end{figure}

Measurements of the production of multi-strange baryons $\Xi \to \Lambda +
\pi \to p + \pi + \pi$ and $\Omega \to \Lambda + K \to p + \pi + K$ were also reported at this conference and in~\cite{ref:AntoninSQM}. The $\Xi^{-}$ and $\Omega^{-}$ baryon and their antiparticle spectra are shown in Figure~\ref{fig:multipp} (left, top panel) with the corresponding L\'evy-Tsallis fits.  When compared to Monte Carlo 
event generators multi-strange production is under-predicted by various PYTHIA
tunes, though the most-recent Perugia-2011 tune shows
an overall better agreement with the data reported in Figure~\ref{fig:multipp} (left, bottom panel). The $p_{t}$-integrated $dN/dy$ and $\langle p_{t} \rangle$ are reported in Figure~\ref{fig:multipp} (right) as a function of the collision energy together with STAR~\cite{ref:multistar} and CMS~\cite{ref:multicms} results. They are observed to increase with collision energy and due to the precision of the ALICE measurements a significant separation between the $\langle p_{t} \rangle$ of $\Xi^{\pm}$ and $\Omega^{\pm}$ is observed in proton-proton collisions at 7 TeV.

\section{Results in Pb--Pb collisions}\label{sec:pbpbresults}

\begin{figure}[t]
  \centering
  \begin{minipage}[c]{0.48\linewidth}
    \centering
    \includegraphics[viewport=10 0 517 383, width=\textwidth]{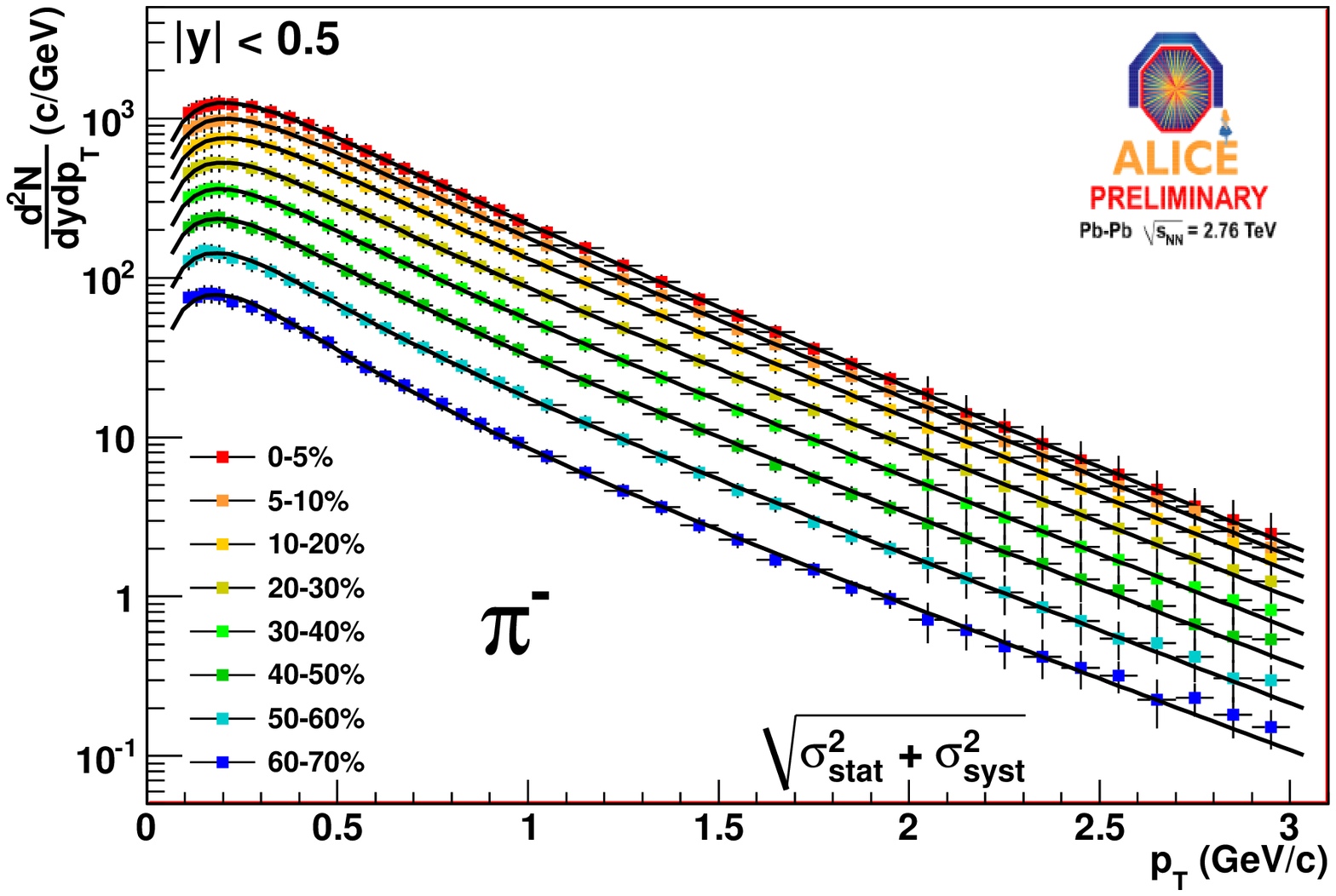}
  \end{minipage}
  \hfill
  \begin{minipage}[c]{0.48\linewidth}
    \centering
    \includegraphics[viewport=10 0 517 383, width=\textwidth]{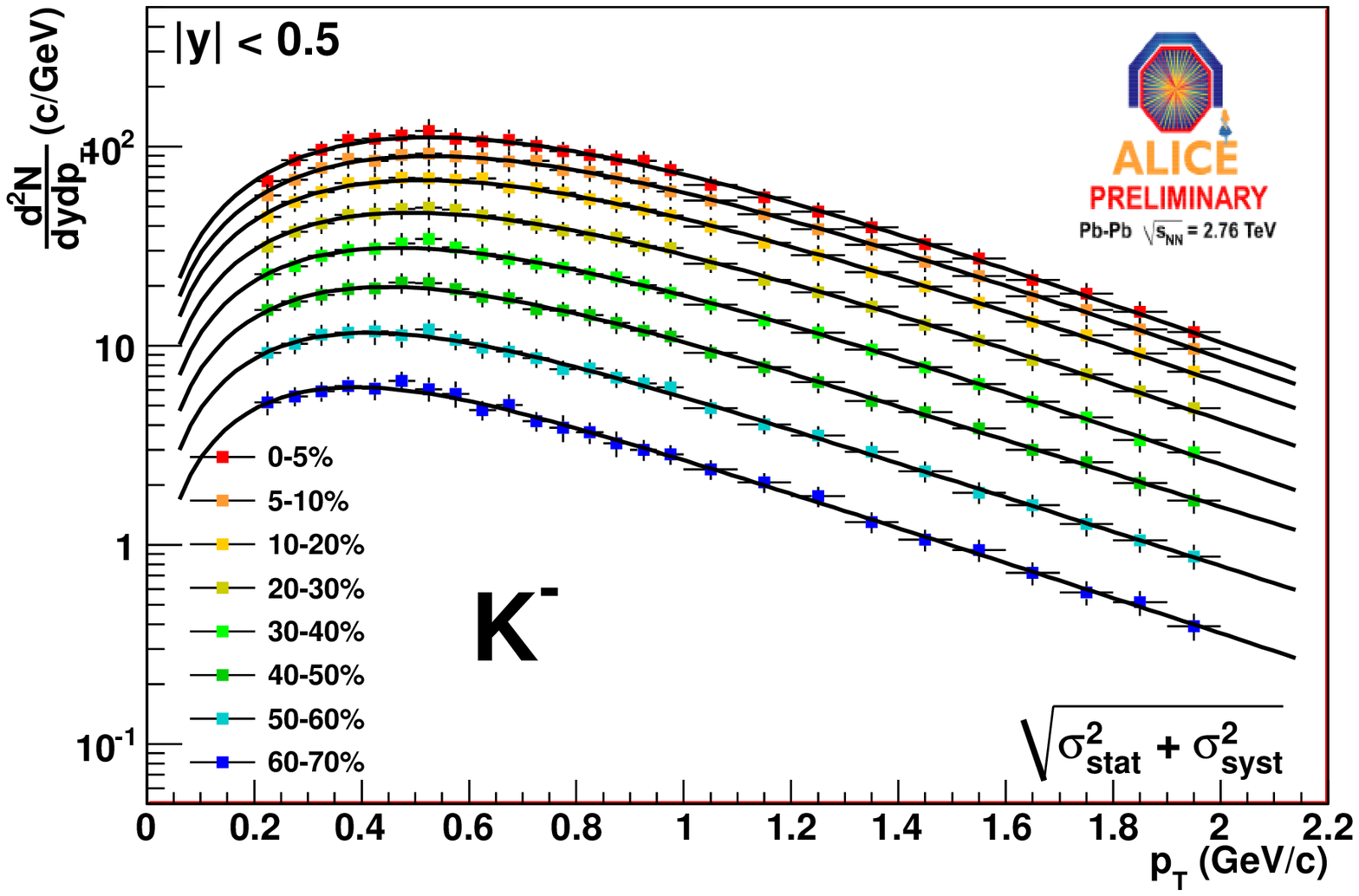}
  \end{minipage}

  \begin{minipage}[c]{0.48\linewidth}
    \centering
    \includegraphics[viewport=10 0 517 383, width=\textwidth]{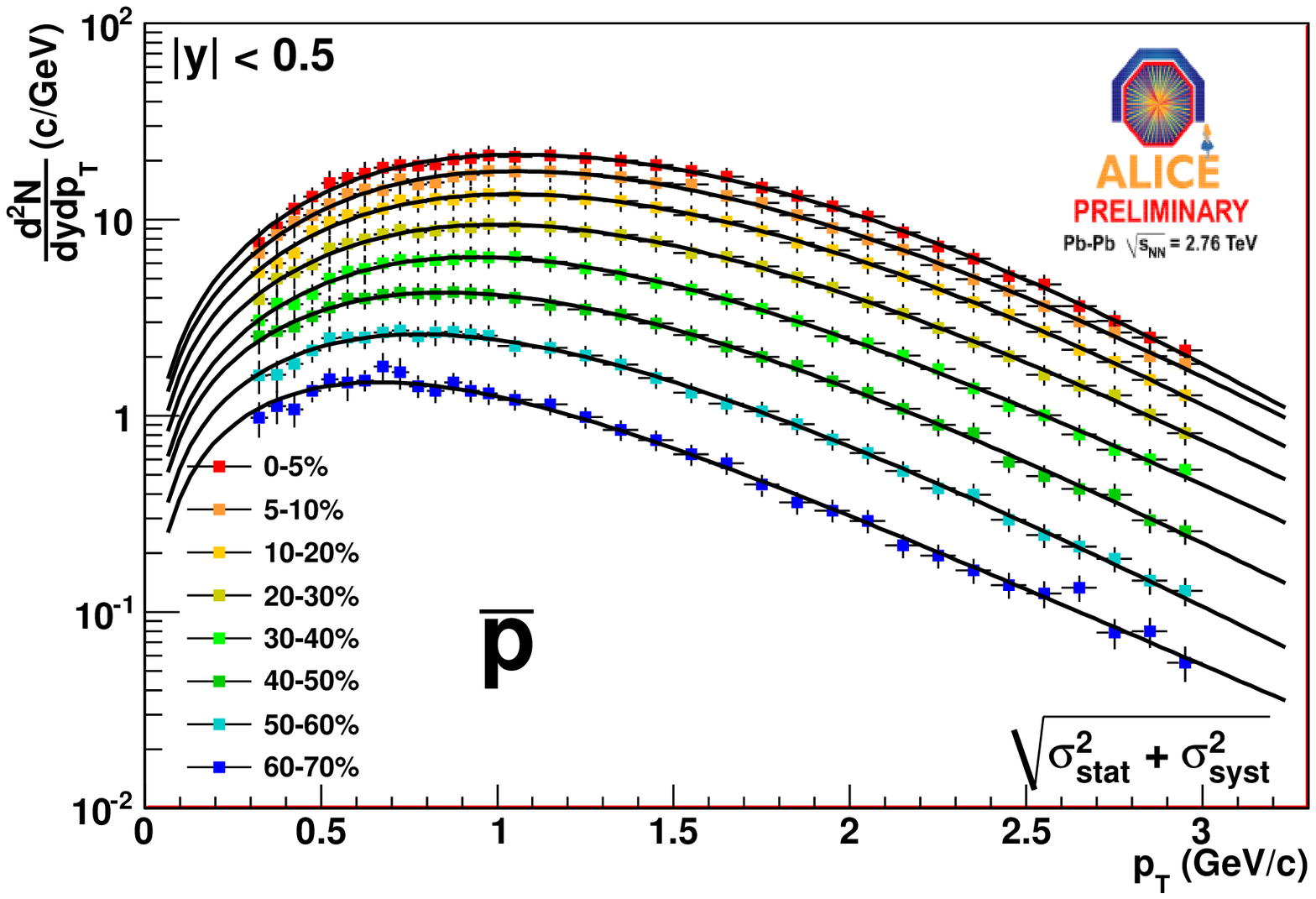}
  \end{minipage}
  \hfill
  \begin{minipage}[c]{0.48\linewidth}
    \centering
    \includegraphics[viewport=10 0 517 383, width=\linewidth]{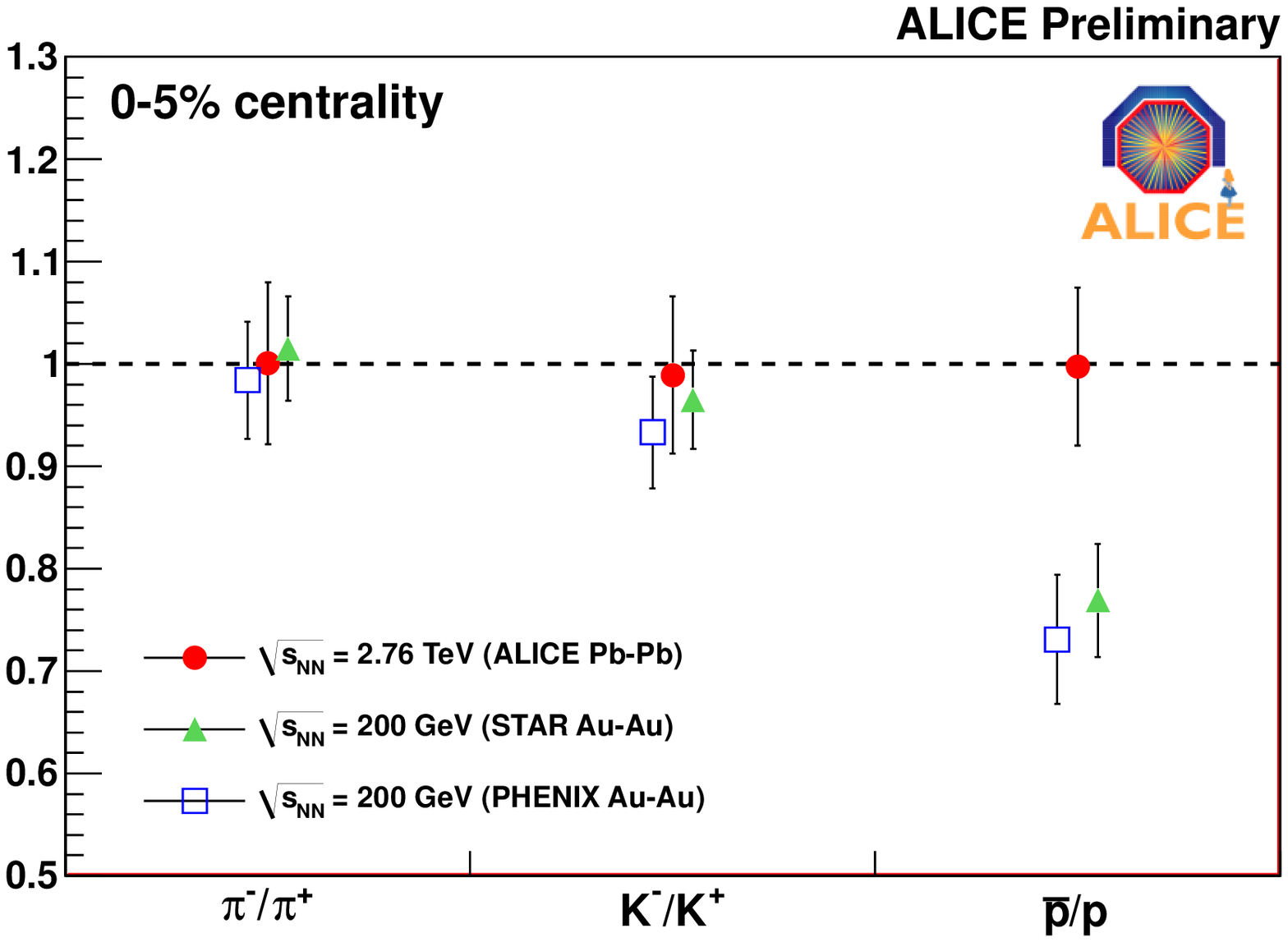}
  \end{minipage}
  \caption{Transverse momentum spectra of primary $\pi^{-}$ (top left),
    $K^{-}$ (top right),
      $\bar{p}$ (bottom left) and corresponding fits in Pb--Pb collisions at
      $\sqrt{s_{\rm NN}}$~=~2.76~TeV. Antiparticle/particle production ratios in the 0-5\% most central
    collisions (bottom right).}
  \label{fig:pbpbspectra}
\end{figure}

Hadron spectra are
measured in several centrality classes (see~\cite{ref:ALICEpbpb} for details
on centrality selection) from 100~MeV/c up to 3~GeV/c for pions, from
200~MeV/c up to 2~GeV/c for kaons and from 300~MeV/c up to 3~GeV/c for
protons and antiprotons. Individual fits to the data are performed following a
blast-wave parameterization~\cite{ref:blastwave} to extrapolate the
spectra outside the measured $p_{T}$ range. The measured spectra and corresponding fits are shown in
Figure~\ref{fig:pbpbspectra} for primary $\pi^{-}$ (top left), $K^{-}$ (top
right) and $\bar{p}$ (bottom left). Average transverse
momenta $\langle p_{T} \rangle$ and integrated 
production yields $dN/dy$ are obtained using the measured data points and the
extrapolation. Antiparticle/particle integrated production ratios are observed to be
consistent with unity for all hadron species in all centrality classes suggesting
that the baryo-chemical potential $\mu_{B}$ is close to zero as expected at LHC
energies. Figure~\ref{fig:pbpbspectra} (bottom right) compares ALICE results with
RHIC Au--Au 
collisions at $\sqrt{s_{\rm NN}}$ = 200 GeV~\cite{ref:RHIC} for the 0-5\%
most central collisions. 

\begin{figure}[t]
  \begin{minipage}[c]{0.48\linewidth}
    \centering
    \includegraphics[viewport=10 0 517 383, width=\textwidth]{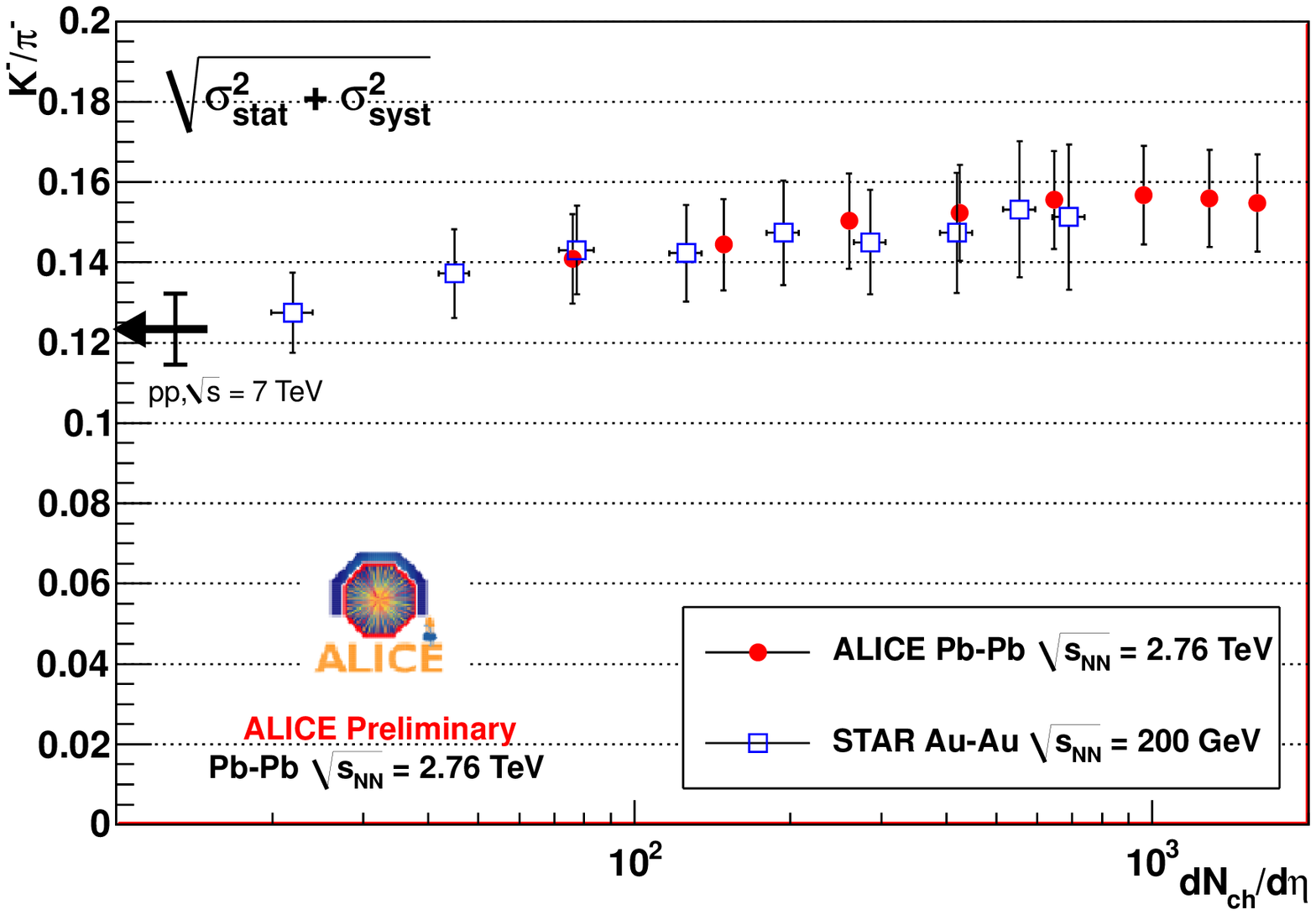}
  \end{minipage}
  \hfill
  \begin{minipage}[c]{0.48\linewidth}
    \centering
    \includegraphics[viewport=10 0 517 383, width=\textwidth]{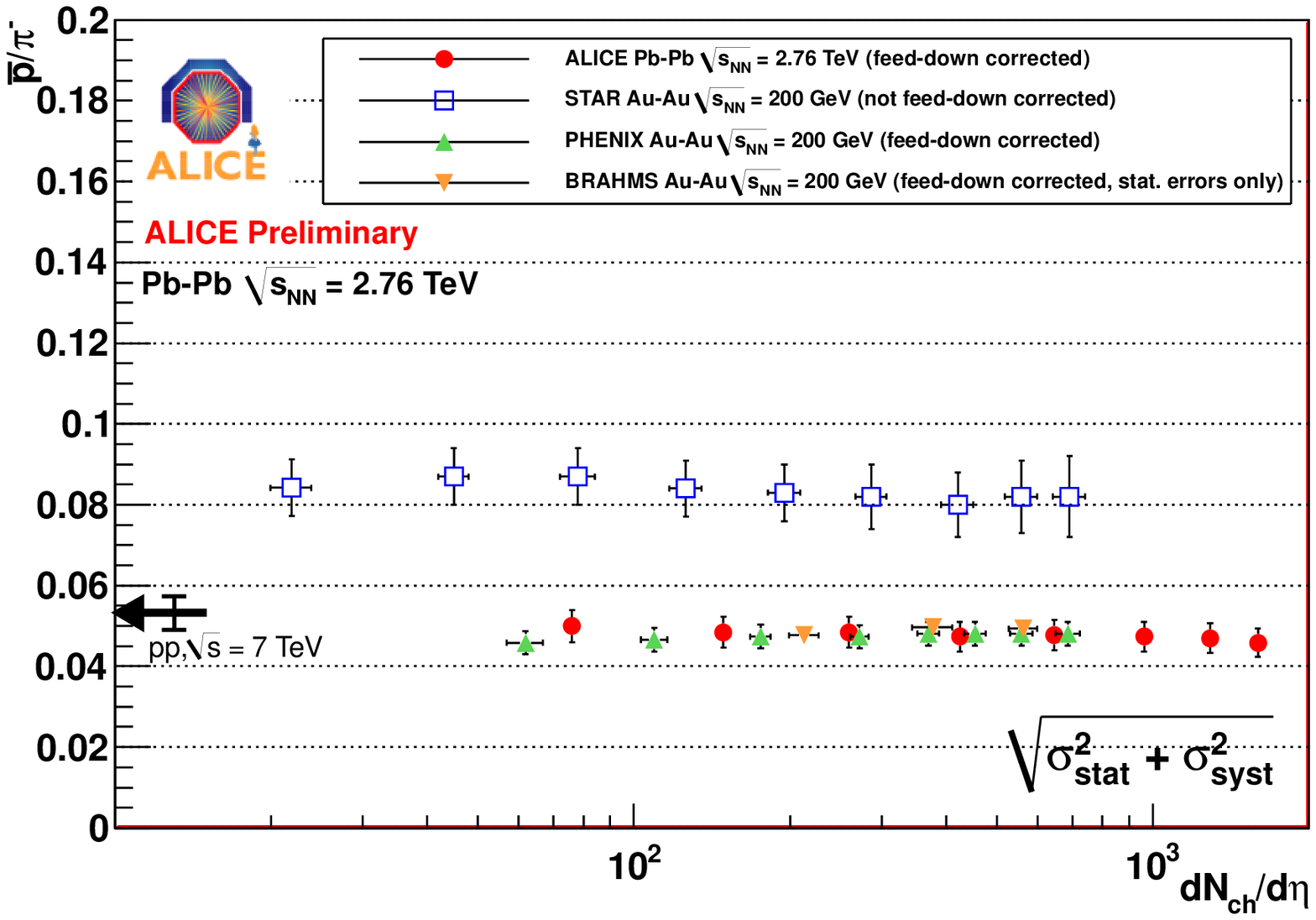}
  \end{minipage}
  \caption{$K^{-}/\pi^{-}$ (left) and $\bar{p}/\pi^{-}$ (right) production ratios as a function of
    $dN_{ch}/d\eta$ in Pb--Pb collisions at $\sqrt{s_{\rm NN}}$~=~2.76~TeV compared to RHIC data.}
  \label{fig:kapiprpiratio}
\end{figure}

The $p_{T}$-integrated $K^{-}/\pi^{-}$ and $\bar{p}/\pi^{-}$ ratios are shown in
Figure~\ref{fig:kapiprpiratio} as a function of the charged-particle
density $dN_{ch}/d\eta$~\cite{ref:ALICEpbpb} and are compared with RHIC data at 
$\sqrt{s_{\rm NN}}$~=~200~GeV and
ALICE proton-proton results at
$\sqrt{s}$~=~7~TeV. $K^{-}/\pi^{-}$ production nicely 
follows the trend measured by STAR. $\bar{p}/\pi^{-}$ results are similar to
previous measurements performed by PHENIX and BRAHMS (proton measurements reported by STAR are inclusive). Finally, the $\bar{p}/\pi^{-}$ ratio measured at the LHC ($\sim$~0.05) is
significantly lower that the value expected from statistical model predictions
($\sim$~0.07-0.09) with a chemical freeze-out temperature of $T_{ch} =
160-170$~MeV at the LHC~\cite{ref:statmodels}. 

\begin{figure}[t]
  \centering
  \begin{minipage}[c]{0.48\linewidth}
    \centering
    \includegraphics[viewport=10 0 517 383, width=\textwidth]{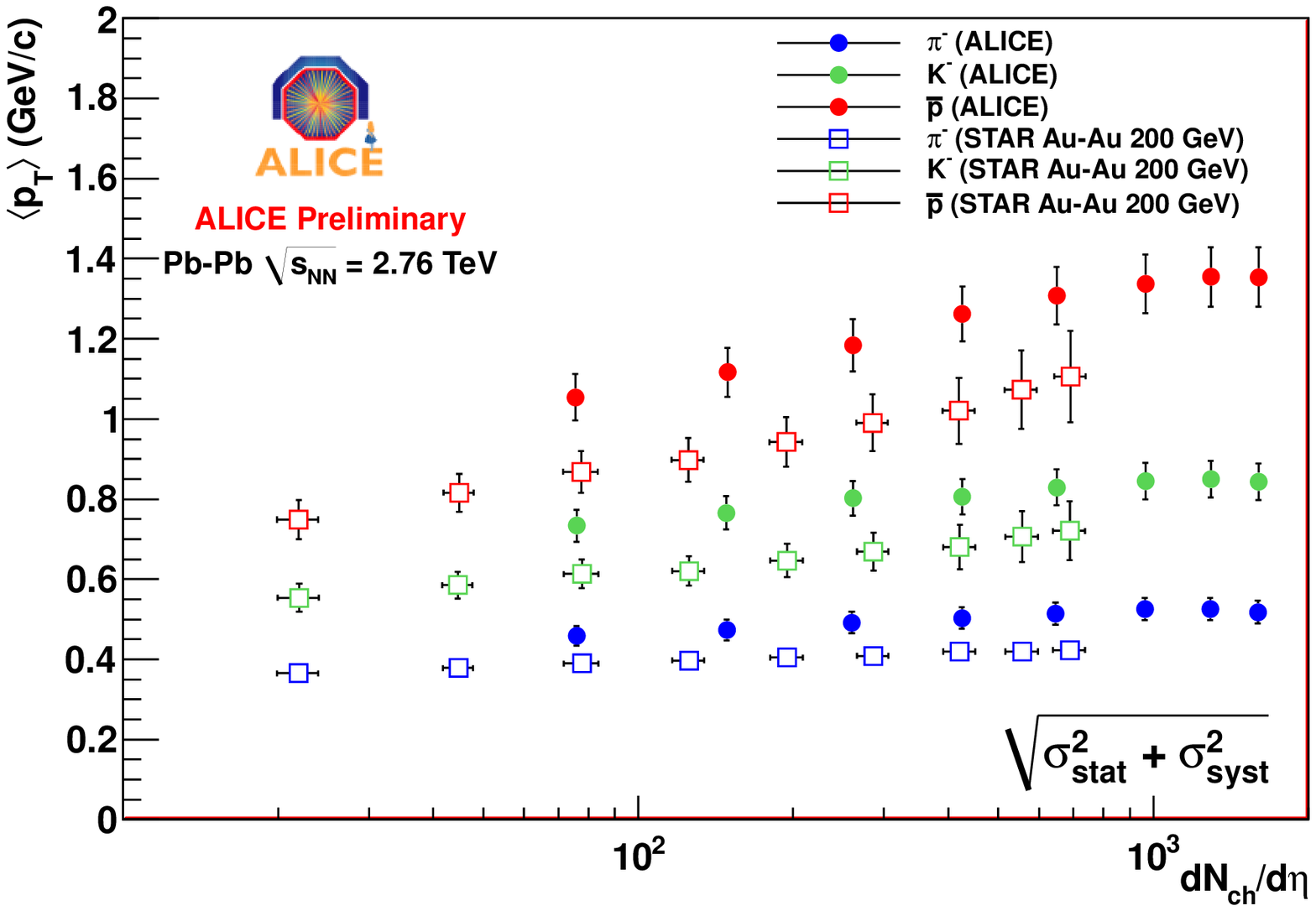}
  \end{minipage}
  \hfill
  \begin{minipage}[c]{0.48\linewidth}
    \centering
    \includegraphics[viewport=10 0 517 354, width=\textwidth]{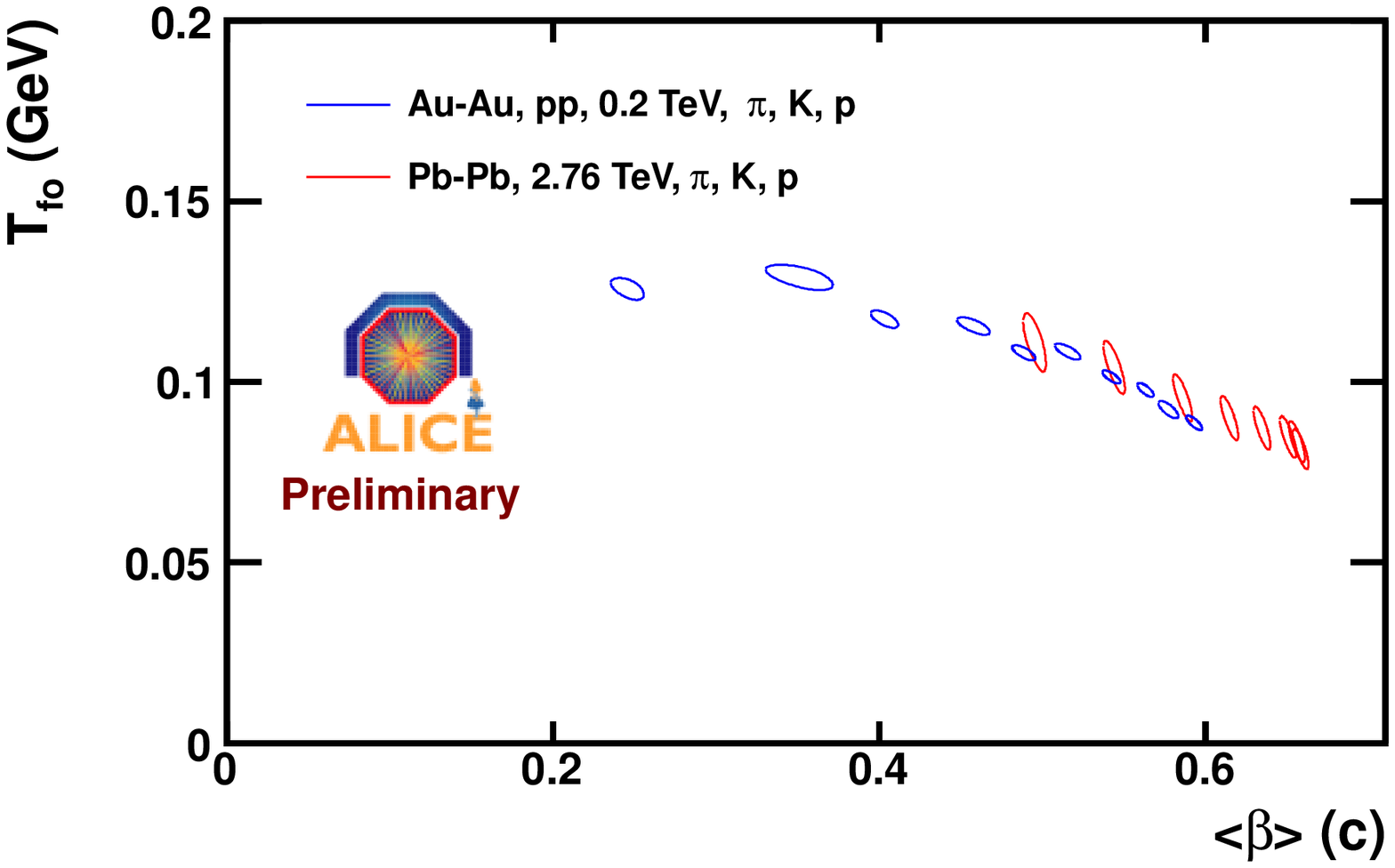}
  \end{minipage}
    \caption{Hadron $\langle p_{T} \rangle$ as a function of the
      charged-particle density $dN_{ch}/d\eta$ in Pb--Pb collisions at
      $\sqrt{s_{\rm NN}}$~=~2.76~TeV (left). Thermal freeze-out parameters $T_{fo}$ and $\langle \beta \rangle$
      from combined blast-wave fits compared to RHIC data (right).}
    \label{fig:meanptblastwave}
\end{figure}

\begin{figure}[t]
  \centering
  \begin{minipage}[c]{0.48\linewidth}
    \centering
    \includegraphics[width=0.9\textwidth]{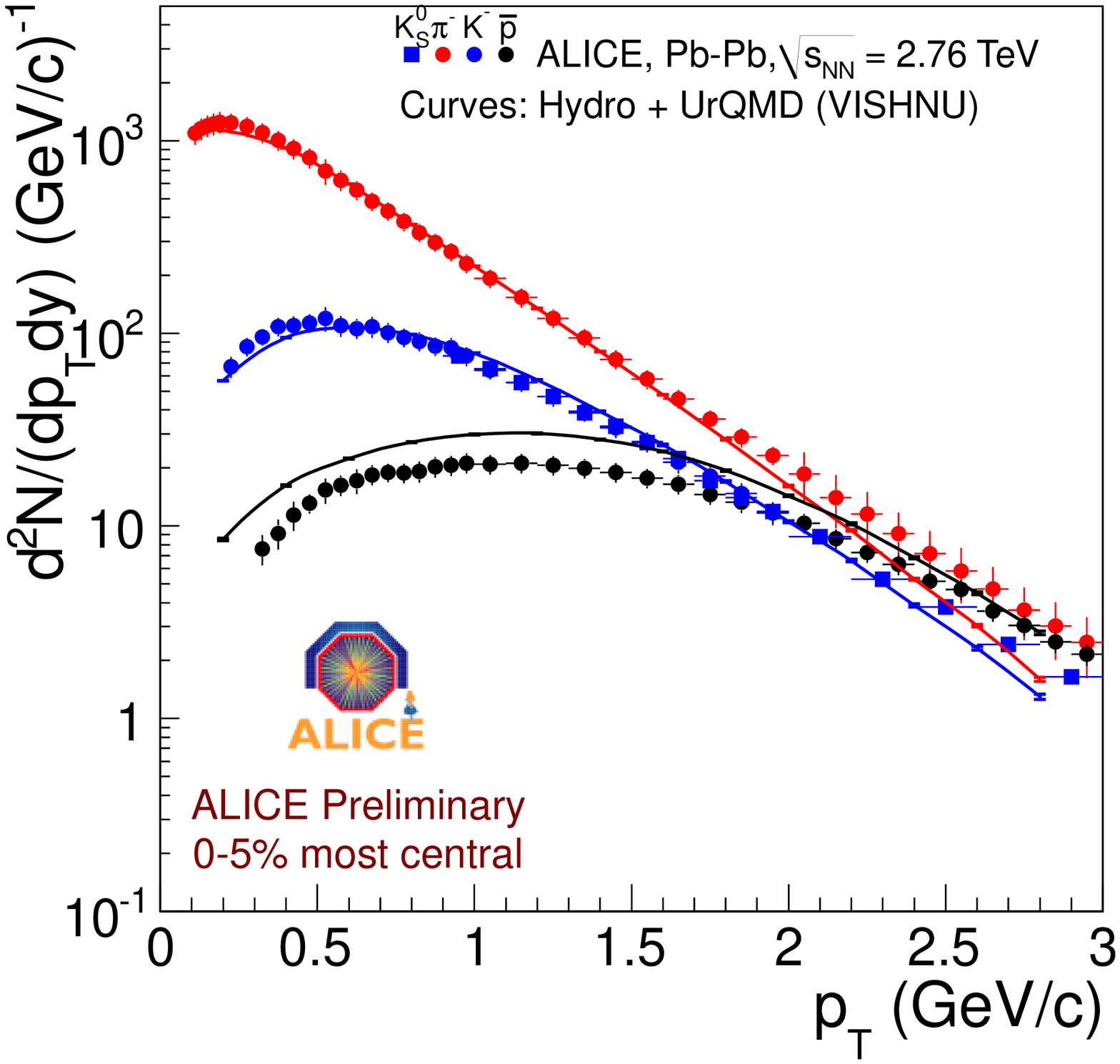}
  \end{minipage}
  \hfill
  \begin{minipage}[c]{0.48\linewidth}
    \centering
    \includegraphics[viewport=50 0 651 482, width=\textwidth]{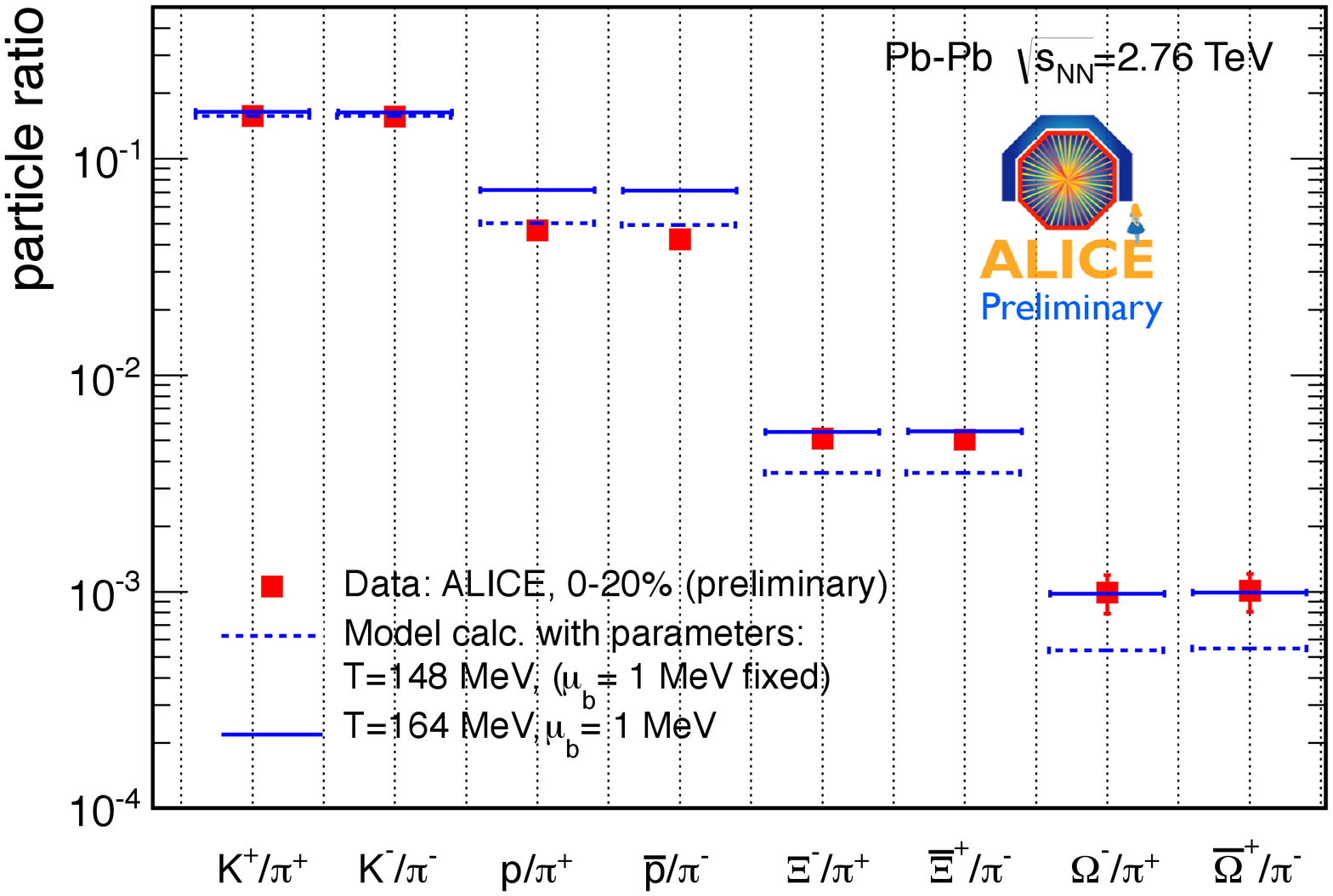}
  \end{minipage}
  \caption{$\pi^{-}$, $K^{-}$, $\bar{p}$ spectra in 0-5\% central Pb--Pb
    collisions compared to hydrodynamic model predictions~\cite{ref:vishnu,ref:vishnuspectra} (left). Hadron-production ratios compared to thermal model predictions~\cite{ref:statmodels} (right).}
    \label{fig:vishnuthermal}
\end{figure}

The measured hadron $\langle p_{T} \rangle$'s are shown in
Figure~\ref{fig:meanptblastwave} (left) as a function of $dN_{ch}/d\eta$ for $\pi^{-}$, $K^{-}$ and $\bar{p}$ and
are compared to STAR results in Au--Au collisions at
$\sqrt{s_{\rm NN}}$~=~200~GeV. The measured spectra at the LHC are observed to be
harder than at RHIC for similar $dN_{ch}/d\eta$. A detailed study of the
spectral shapes has been done in order to give a quantitative estimate of
the thermal freeze-out temperature $T_{fo}$ and the average transverse flow
$\langle \beta \rangle$. A combined blast-wave fit of the spectra has
been performed in the ranges 0.3-1.0~GeV/c, 0.2-1.5~GeV/c and 0.3-3.0~GeV/c
for pions, kaons and protons respectively. While the $T_{fo}$ parameter is
slightly sensitive to the pion fit range because of feed-down of
resonances the transverse flow $\langle \beta \rangle$ measurement is not,
being dominated by the proton spectral shape. The results obtained on the
thermal freeze-out properties in different centrality bins are compared with
similar measurements performed by the STAR Collaboration at lower energies in
Figure~\ref{fig:meanptblastwave} (right). A stronger radial flow is observed with respect to
RHIC, being about 10\% larger in the most central collisions at the LHC. The data are also compared to predictions from hydrodynamic models. As already
reported in~\cite{ref:MicheleQM} the pure hydrodynamic
predictions~\cite{ref:purehydro} cannot reproduce the proton shape. A similar
disagreement was observed when comparing proton elliptic flow $v_{2}$ to the
same model~\cite{ref:SnellingsQM}. A new
calculation has been performed by U.Heinz \emph{et al.} using a hybrid 
model which adds an hadronic rescattering and freeze-out stage to the pure
viscous dynamics~\cite{ref:vishnu}. These new predictions~\cite{ref:vishnuspectra} are compared to the data in
Figure~\ref{fig:vishnuthermal} (left) and the agreement with the proton shape
is better than a pure hydrodynamic picture. This
suggests that extra flow builds up in the hadronic phase. The
difference in the proton yield can be ascribed to the fact that the model
derives yields from a thermal model with $T_{ch} = 165$~MeV. It is worth that this model also
reproduces the shape of elliptic flow of identified 
hadrons as already reported in~\cite{ref:NoferoSQM}.

\begin{figure}[t]
  \centering
  \begin{minipage}[c]{0.48\linewidth}
    \centering
    \includegraphics[viewport=10 0 517 383, width=0.9\textwidth]{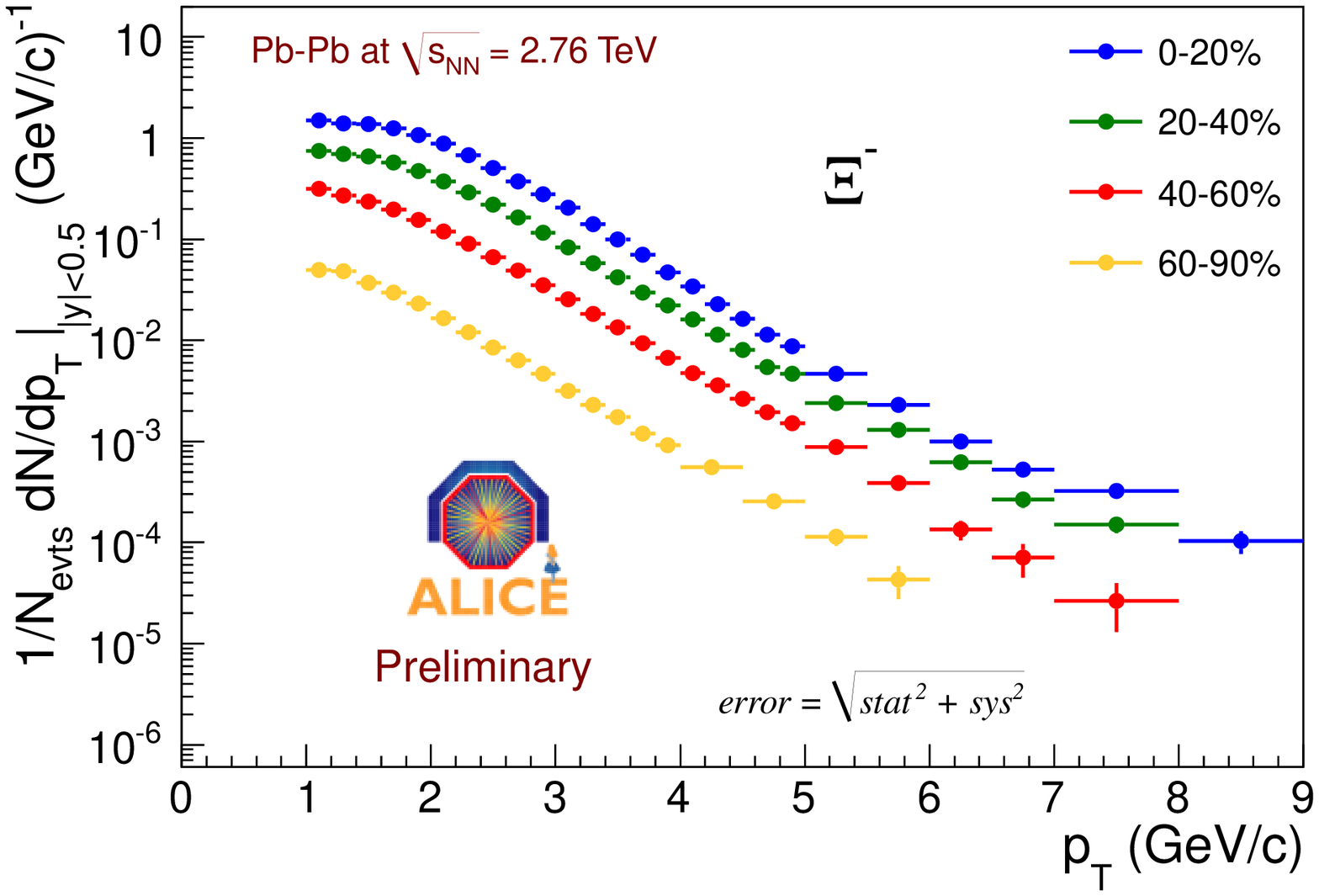}
  \end{minipage}
  \hfill
  \begin{minipage}[c]{0.48\linewidth}
    \centering
    \includegraphics[width=\textwidth]{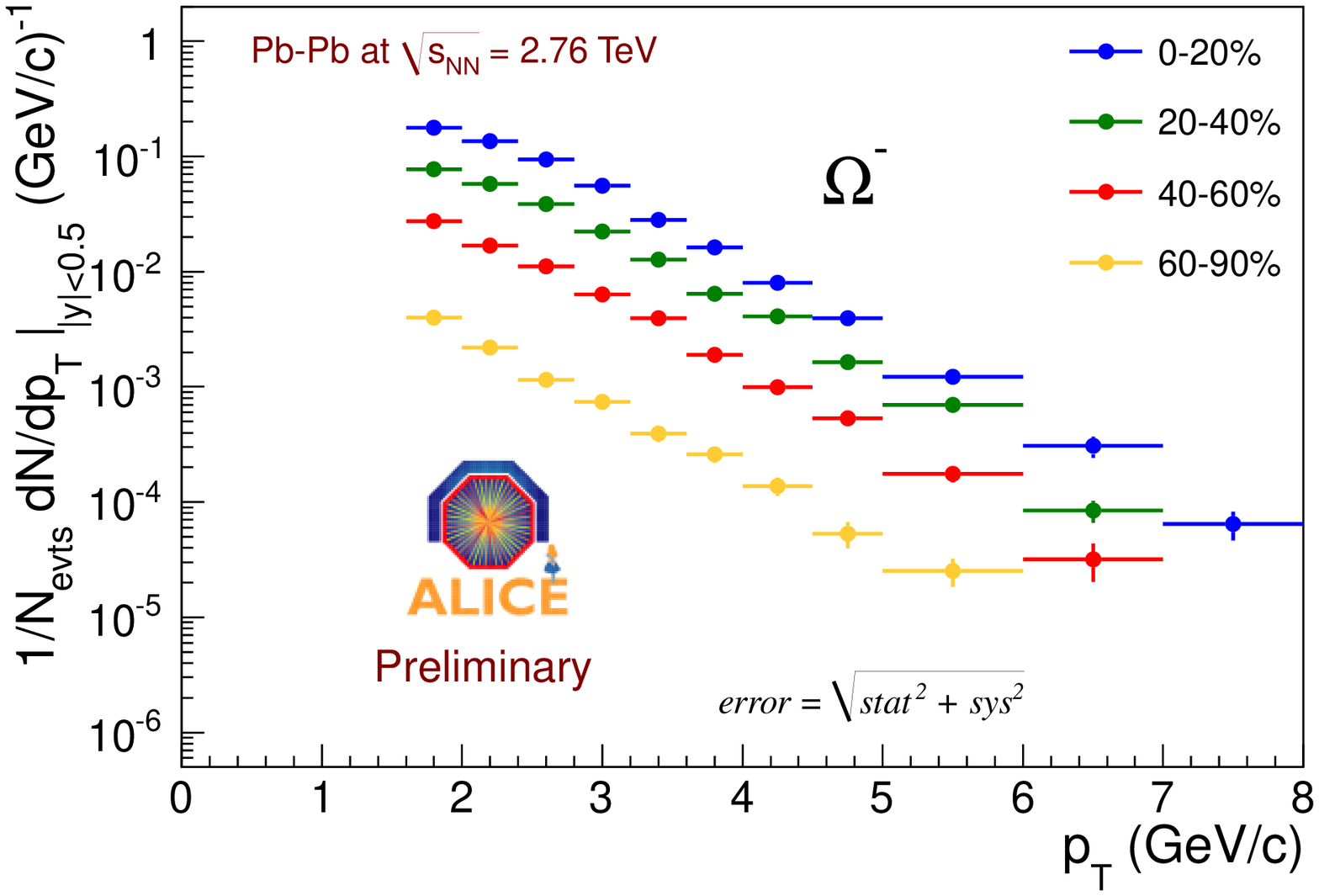}
  \end{minipage}
  \caption{Transverse momentum spectra of $\Xi^{-}$ (left) and $\Omega^{-}$ (right) measured in four centrality classes in Pb--Pb collisions at $\sqrt{s_{\rm NN}}$~=~2.76~TeV.}
    \label{fig:multipbpb}
\end{figure}

Finally, the production of multi-strange baryons in Pb--Pb
collisions has also been reported at this conference and in~\cite{ref:MariellaSQM}. The transverse momentum spectra of $\Xi^{-}$ and $\Omega^{-}$ measured in several centrality classes are shown in Figure~\ref{fig:multipbpb}. The measured yields of
several hadrons normalized to the pion yield in 0-5\% central collisions are
compared to thermal model predictions in Figure~\ref{fig:vishnuthermal}
(right). With a temperature of $T_{ch} = 164$~MeV  predicted by A. Andronic
\emph{et al.}~\cite{ref:statmodels} the model reproduces both kaon and multi-strange
production but fails with protons. The same model can be tuned to reproduce
proton yields with an ad-hoc $T_{ch} = 148$~MeV, though multi-strange
production is underestimated in this case.

\section{Conclusions}\label{sec:conclusions}
The transverse momentum spectra of $\pi^{\pm}$, $K^{\pm}$, $p$ and $\bar{p}$ have
been measured with ALICE in pp collisions at $\sqrt{s}$~=~900~GeV and
7~TeV and in Pb--Pb collisions at $\sqrt{s_{\rm NN}}$~=~2.76~TeV,
demonstrating the excellent PID capabilities of the
experiment. Proton-proton results show no evident $\sqrt{s}$ dependence in
hadron production ratios. Currently available Monte Carlo
generators and tunes cannot simultaneously reproduce charge kaon and resonance
production and multi-strange production is generally
underestimated at low-$p_{T}$. In Pb--Pb
collisions $\bar{p}/\pi^{-}$ integrated ratio is significantly 
lower than statistical model predictions with a chemical freeze-out temperature $T_{ch} =
160-170$~MeV. The average transverse momenta and the spectral shapes indicate
a $\sim$10\% stronger radial flow than at RHIC energies.

\small{
  
}

\end{document}